# Confocal laser scanning microscopy: A tool for rapid optical characterization of 2D materials

Vishal Panchal[*1,2], Yanfei Yang[2,3], Guangjun Cheng[2], Jiuning Hu[2], Mattias Kruskopf[2], Chieh-I Liu[2,4], Albert F. Rigosi[2], Christos Melios[1], Angela R. Hight Walker[2], David B. Newell[2], Olga Kazakova[1], and Randolph E. Elmquist[*2]

[1]National Physical Laboratory, Hampton Road, Teddington, TW11 0LW, UK

[2]National Institute of Standards and Technology, Gaithersburg, MD 20899, USA

[3]Joint Quantum Institute, University of Maryland, College Park, MD 20742, USA

[4]National Taiwan University, Taipei, 10617, Taiwan

**ABSTRACT:** Confocal laser scanning microscopy (CLSM) is a non-destructive, highly-efficient optical characterization method for large-area analysis of graphene on different substrates, which can be applied in ambient air, does not require additional sample preparation, and is insusceptible to surface charging and surface contamination. CLSM leverages optical properties of graphene and provides greatly enhanced optical contrast and mapping of thickness down to a single layer. We demonstrate the effectiveness of CLSM by measuring mechanically exfoliated and chemical vapor deposition graphene on $Si/SiO_2$, and epitaxial graphene on SiC. In the case of graphene on $Si/SiO_2$, both CLSM intensity and height mapping is powerful for analysis of 1-5 layers of graphene. For epitaxial graphene on SiC substrates, the CLSM intensity allows us to distinguish features such as dense, parallel 150 nm wide ribbons of graphene (associated with the early stages of the growth process) and large regions covered by the interfacial layer and 1-3 layers of graphene. In both cases, CLSM data shows excellent correlation with conventional optical microscopy, atomic force microscopy, Kelvin probe force microscopy, conductive atomic force microscopy, scanning electron microscopy and Raman mapping, with a greatly reduced acquisition time. We demonstrate that CLSM is an indispensable tool for rapid analysis of mass-produced graphene and is equally relevant to other 2D materials.

Wafer-scale graphene material is of interest for quantum Hall resistance standards[1–5] and future nanoelectronics[6,7], such as high frequency electronics[8–15] and photonics[16,17]. Single-domain epitaxial graphene (EG) grown on the silicon face of SiC(0001)[18] has several advantages, such as removing the need to transfer the graphene onto an insulating substrate for device processing, as is the case for chemical vapor deposition (CVD) growth. Recent progress in CVD and EG growth demonstrates the potential for mass production of homogeneous graphene at the wafer-scale[5,19–22], and increases the demand for a characterization method that is fast, accurate, and accessible. Previously, optical microscopy was demonstrated to be a useful tool for rapid identification of layer inhomogeneities in EG over hundreds of micrometers[23]. However, low contrast and poor spatial resolution are significant limiting factors. Currently, Raman spectroscopy and scanning probe microscopy (SPM), including Kelvin probe force microscopy (KPFM), are the most widely used methods of characterizing the material quality. Raman spectroscopy is a non-destructive tool for structural analysis of graphene, and is furthermore sensitive to the doping level and strain in graphene[24–29]. For 1-layer graphene (LG) the fingerprint in the Raman spectrum is a symmetric 2D peak at ~2700 cm$^{-1}$ that can be fitted by a single Lorentzian[26,30]. However, a careful analysis of the shape of the 2D peak is required to identify increasing number of graphene layers. The topography imaged by atomic force microscopy (AFM) is another method for determining layer thickness of CVD or exfoliated graphene on various substrates. However, for graphene on SiC, identification of EG layers from the topography is far from straightforward using AFM alone[31], due to terrace structure of the SiC substrate, which develops concurrently with the EG and thus has a strong influence on the layer growth and uniformity[5,32]. Recently, KPFM was shown to be a more reliable method for distinguishing the number of EG layers[33]. Nonetheless, Raman and SPM methods are time consuming and typically limit the scan size to a few tens of micrometers. Scaling up the production process requires fast and accurate characterization of the material quality at the wafer scale, while at the same time retaining sub-micrometer spatial resolution.

In this report, we demonstrate that reflection mode CLSM is a superior tool for rapid characterization of large-area graphene and graphene nanostructures on $Si/SiO_2$ and SiC, compared to conventional optical microscopy (OM), Raman spectroscopy, AFM, conductive atomic force microscopy (C-AFM), KPFM, and scanning electron microscope (SEM) methods. CLSM can simultaneously produce intensity and topography images as well as has a lateral resolution that can be pushed beyond the optical diffraction limit. The depth-of-field is enhanced by digitally selecting in-focus regions from multiple images at different focal planes, enabling high resolution over larger areas. First, we discuss CLSM results on various thicknesses of exfoliated graphene transferred to $Si/SiO_2$ substrate. By comparing the results to AFM and Raman measurements, we present a method for assessing the CLSM intensity and height for graphene of different thicknesses. Next, we apply CLSM and Raman spectroscopy to CVD-grown graphene transferred to $Si/SiO_2$ substrate to demonstrate fast and accurate, large-scale analysis. Finally, we apply CLSM to epitaxial graphene on SiC demonstrating the speed, accuracy, and versatility of CLSM compared to OM, SEM, AFM, KPFM, C-AFM and Raman microscopy.

**RESULTS**

**1. Exfoliated graphene on $Si/SiO_2$**

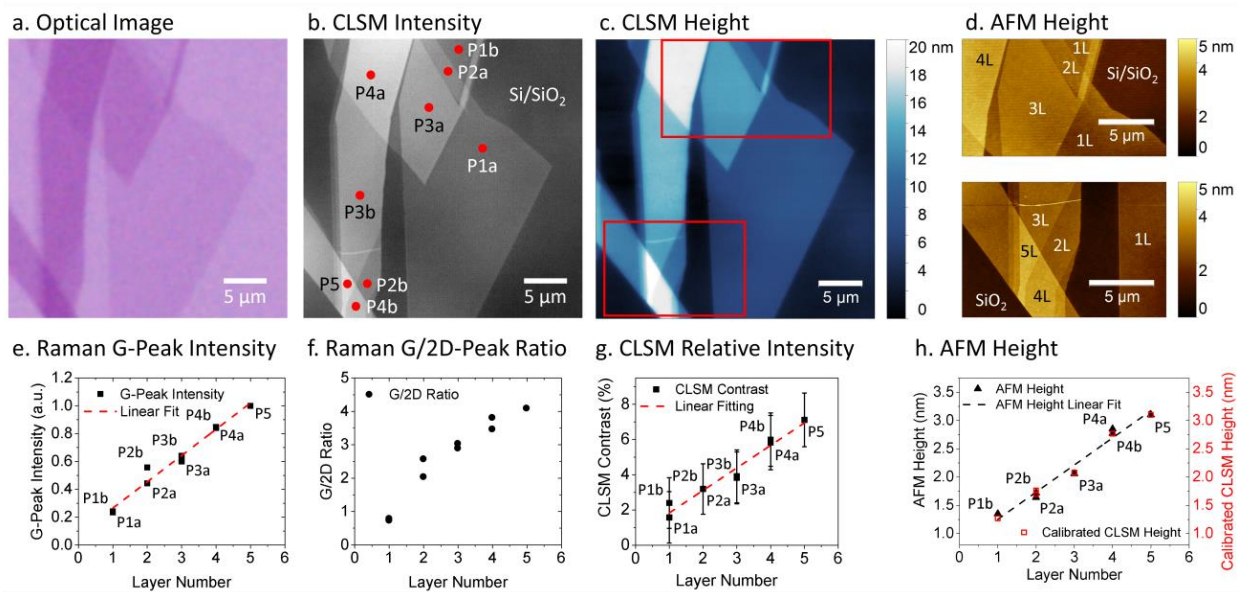

**Figure 1.** Characterization of exfoliated graphene on Si/SiO₂ by CLSM, compared to OM, AFM, and Raman spectroscopy. (a) OM, (b) CLSM intensity and (c) CLSM height images. (d) AFM images of the areas marked in (c). (e) G-peak intensity and (f) G/2D-peak intensity ratio. (g) CLSM relative intensity measured at the red points marked in (b), as a function of the graphene layer thickness. (h) CLSM height measurement calibrated with AFM height measurement for 1-5 graphene layers. Raman data was acquired with 514.5 nm excitation.

To demonstrate the potential of CLSM for the characterization of 2D materials, we studied exfoliated graphene transferred to Si/SiO₂ substrates. The OM images of the sample was carried out on a Nikon L200N optical microscope[34] in the reflection mode using white light. Figure 1a, 1b and 1c are OM, CLSM intensity and CLSM height images, respectively, of exfoliated graphene flakes on Si substrate covered by 300 nm of SiO₂. Both the imaging techniques are performed in reflection mode. Each graphene layer absorbs 2.3% of the incident light[35]. The same region imaged with the CLSM shows significantly higher signal-to-noise ratio than OM as well as provides *in-situ* map of the height. Raman spectra were recorded at the nine different regions of the sample as indicated by red dots in Fig. 1b. The spectra for 1-5 LG are shown in Supplementary Fig. S1c. Analysis of points P1a and P1b match the description of single layer graphene, where their 2D-peaks can be fitted by a single Lorentzian (FWHM of 26.8 cm$^{-1}$ and 28.1 cm$^{-1}$, respectively) and the height ratios of G/2D peaks are ~0.7 (Fig. 1f). Ni *et al.* also reported that the G-peak height increases linearly with the number of layers up to 9 layers[36]. This is in good agreement with data shown in Fig. 1e, except for point P2b, where the unexpected behavior may be due to an overlapping of the laser spot with nearby multilayer graphene domains (P4b and P5). Using the layer numbers determined from the Raman analysis, we found that the CLSM relative intensity (compared to Si/SiO$_2$ substrate) also increases approximately linearly with the layer number (Fig. 1g).

The thickness of the first graphene layer (P1b) as measured by AFM is 1.35 ± 0.1 nm (Supplementary Table S1), with subsequent layers being measured as 0.48 nm thick (as estimated from the slope of the linear fitting of the raw AFM data in Supplementary Fig. S2). The larger thickness of the first and subsequent graphene layer(s) can be the result of contamination present between the graphene and substrate interfaces[37]. Although the resolution for CLSM height map is ~10 nm (as specified by the manufacturer), Fig. 1c shows that CLSM is able to distinguish the height of a single layer of graphene. The raw CLSM height values are summarized in Supplementary Table S1, where the first graphene layer is 4.72 ± 0.1 nm, with each subsequent layers being 3.68 nm thick (as estimated from the slope of the linear fitting of the raw CLSM data in Supplementary Fig. S2). Therefore, with proper correction, CLSM can be a fast and reliable method to identify the layer number of exfoliated graphene flakes (Fig. 1h). This is evident by the linearity of the CLSM and theoretical height values.

## 2. CVD graphene on Si/SiO$_2$

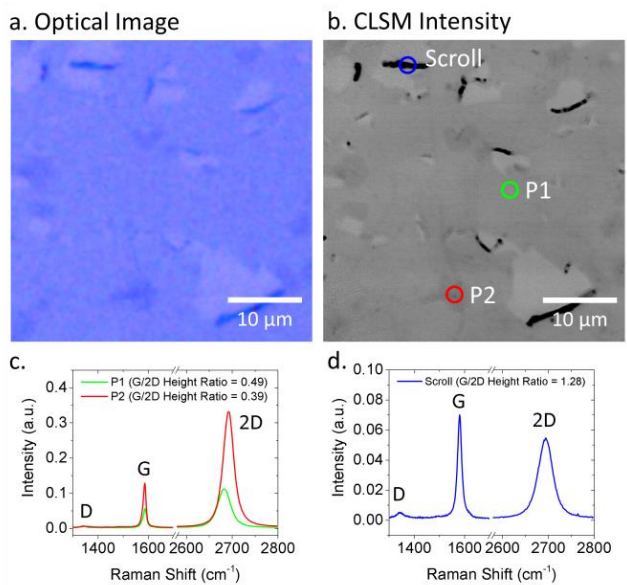

**Figure 2.** OM and CLSM imaging of CVD graphene transferred to Si/SiO$_2$ substrate. (a) OM and (b) CLSM intensity images of CVD-grown graphene transferred to Si/SiO$_2$ substrate, showing tears and scrolls. (c) Raman spectra for points P1 and P2 in (b). (d) Raman spectrum for graphene scroll indicated by a blue circle in (b). Raman data was acquired with 514.5 nm excitation.

Figures 2a and 2b show the OM and CLSM images, respectively, of the CVD graphene that has been transferred to Si/SiO$_2$ substrate. The CLSM image reveals finer structures that are not clearly visible in the OM image, such as the wrinkle at point labeled as P2 in Fig. 2b, which is about 400 nm wide. Although the Raman spectra from both points P1 and P2 display the features characteristic for single layer graphene, there is a blue shift for the G-peak and a red shift for the 2D-peak at P2, which can be attributed to strain from the wrinkle. The Raman spectra of the scroll shows a more pronounced D-peak due to the curvature effect[38]. Furthermore, the graphene layers in the scroll are not tightly packed, which broadens the 2D-peak. The G/2D peak ratio is also larger due to the presence of several layers of graphene within the scroll.

## 3. Epitaxial graphene nanoribbons on SiC

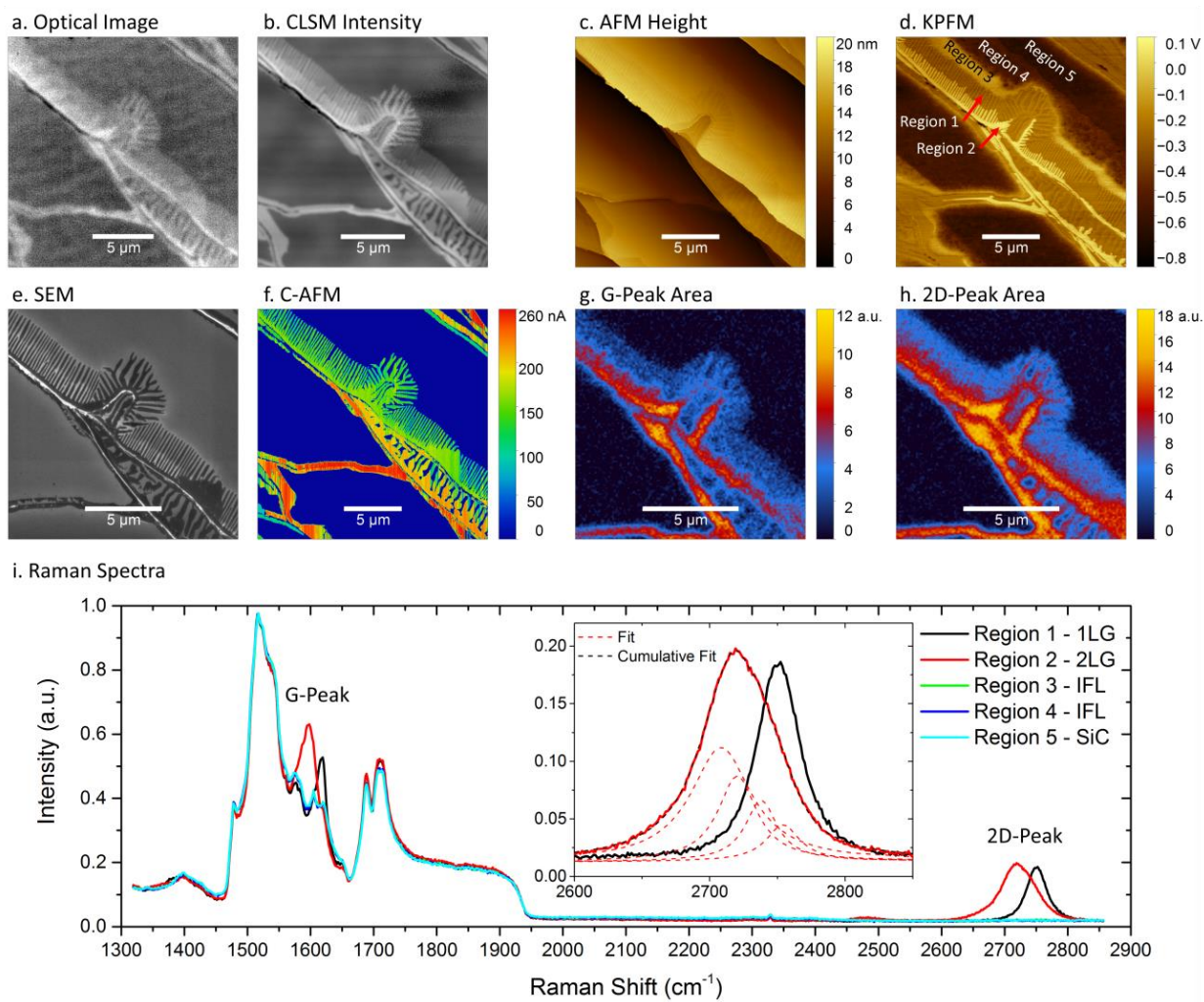

**Figure 3.** Graphene nanoribbons on SiC characterized by various methods. (a) OM, (b) CLSM intensity, (c) AFM, (d) KPFM and (e) SEM with InLens detector, (f) C-AFM images. (g-h) Raman maps obtained for the same area as (a-f), showing the area under the (g) G-peak and (h) 2D-peak. (i) Representative Raman spectra for the five regions indicated in (d), where each is an average of 64 individual spectra from the mapped data. Raman data was acquired with 532 nm excitation.

We first present the epitaxial graphene sample with partial coverage of graphene, as shown in Fig. 3. The sample is comprised of dense 2D nanoribbons with brighter contrast that are not fully resolved by OM (Fig. 3a). Owing to the transparent nature of the SiC substrate, OM imaging is also possible in transmission mode for this particular type of sample (Supplementary Fig. S3), where the contrast is inverted, and the darker regions are associated to graphene with each layer absorbing 2.3% of the incident light[35].

The darker regions in OM and CLSM intensity images are the electrically insulating interfacial layer (IFL) or bare SiC, as verified with a several SPM and Raman techniques (Figs. 3c-g). Imaging in the differential interference contrast (DIC) mode produces a 3-dimensional visualization of the surface morphology with some sacrifice of the graphene contrast (Supplementary Fig. S3c). From comparisons of CLSM, DIC and scanning probe images such as Figs. 3c-d, we find that the graphene nanoribbons preferentially grow from the step edge towards the adjacent upper terrace. The formation of parallel nanoribbons on SiC(0001) terraces has been attributed to diffusion-limited growth[25] that may accompany the decomposition of single, SiC atomic layers.

The CLSM intensity image of Fig. 3b shows the remarkable level of detail produced with an acquisition time of only 10 seconds. Compared to the OM image in reflection mode (Fig. 3a), CLSM provides greatly enhanced lateral resolution and contrast due to the point illumination by a laser source (405 nm wavelength) and by removing the out-of-focus background light with a pinhole at the conjugate focal plane in front of the sensor (i.e., typical confocal configuration). The CLSM intensity image not only shows much higher spatial resolution and fully resolved graphene nanoribbons, but also clearly reveals thin stripes (labeled by red arrows in Fig. 3d) and patches of higher reflectivity along the step edges, indicating two-layer graphene (2LG). Moreover, we determine the lateral resolution of the CLSM to be 150 nm by analyzing the edge spread function from EG images (Fig. S4 and S5).

We obtained additional measurements with AFM, KPFM, SEM, C-AFM, and Raman microscopy, on epitaxial graphene nanoribbons for accurate identification of their structure. The CLSM intensity image in Fig. 3b reveals a dense row of graphene nanoribbons formed from 1LG and 2LG. The KPFM map obtained *in-situ* with the topography (Fig. 3c) reveals substantial variations in the surface potential, with five clear distinct regions (Fig. 3d). Region 1 is assigned to the 1LG nanoribbons which vary in width from ~100 nm to 300 nm and lengths of up to a few micrometers. Region 2 is designated as 2LG nanoribbons, which are similar in width to 1LG but have shorter lengths. Regions 3 and 4 are designated to IFL given that there are no topographical features between them (Fig. 3c), however the surface potential shows significant differences in the charge between the two regions and is attributed to the close proximity of region 3 to the 1LG nanoribbons. Region 5 is designated as SiC, which is least affected by charge from nearby IFL and graphene.

When SEM imaging is carried out with an InLens detector, the parts of the sample with higher work function lead to a stronger suppression of the backscattered electrons from the surface, resulting in darker contrast in SEM image due to a lower electron intensity sensed by the detector[39]. The SEM is able to clearly distinguish graphene regions from IFL with high spatial resolution (Fig 3e), but the only indication of 2LG ribbons is faint haloing of IFL regions surrounding them. Additionally, there is no differentiation between IFL and SiC. The C-AFM map is consistent with the designation of regions 3, 4 and 5 where there was no conduction for these electrically insulating regions (Fig. 3f).

Raman spectra of the same region were also collected and fitted to create maps of the G- and 2D-peak area (Figs. 3g-h), position, width and intensity (Supplementary Fig. S6). Figure 3i shows the average of 64 representative Raman spectra for each of the five regions indicated in Fig. 3d. The signature single Lorentzian 2D-peak for 1LG is observed for region 1, but broadens significantly and forms the signature shoulder for *ab*-stacked 2LG in region 2 (see fitting of the 2D-peak in the inset of Fig. 3i). These maps clearly show that the coverage of graphene is in excellent agreement with CLSM, AFM, KPFM and C-AFM.

The 2D forest of graphene nanoribbons is formed in EG samples produced by face-to-graphite growth with reduced process temperatures or reduced growth times (Supplementary Fig. S7a). In this case, the dense 2D graphene nanoribbon forest along with its conspicuous optical contrast to the IFL patches is a characteristic of incomplete EG coverage. The graphene nanoribbons will eventually merge to form continuous graphene in a succeeding growth (shown in Supplementary Fig. S7b), and the CLSM contrast features will evolve accordingly.

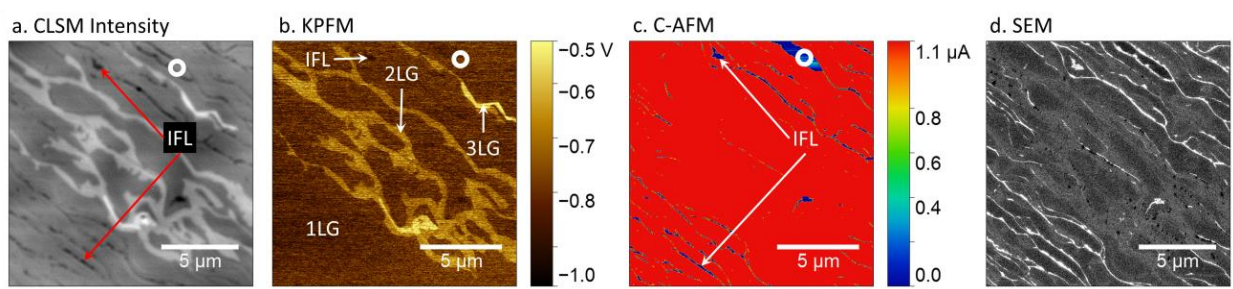

**Figure 4.** Epitaxial graphene on SiC showing IFL, 1LG, 2LG and 3LG. (a) CLSM, (b) KPFM, (c) C-AFM and (d) SEM image with InLens detector.

Next, we investigate samples with full coverage of epitaxial graphene using CLSM, KPFM, C-AFM and SEM (Figs. 4a-d, respectively). The CLSM intensity image shows that the sample is predominantly covered by 1LG (as conformed by Raman spectra; not shown) and about 10% of the area is covered by narrow patches of bi- (2LG) and tri-layer graphene (3LG) domains (brightest contrast). The higher intensity of the reflected light from thicker graphene layer in the CLSM image is consistent with the linear increase of reflectivity reported by Ivanov *et al*., where they measured the power of reflected laser light from the surface of EG on SiC[40]. Comparison of Fig. 4a to the surface potential map (Fig. 4b) of the same region confirms the designation of 1LG, 2LG, and 3LG. However, the darkest lines and patches in Fig. 4a indicated by the red arrows are not as clearly apparent in the KPFM image due to the proximity effect from charging. These features are confirmed to be insulating IFL or SiC by C-AFM, where zero current is measured (Fig. 4c). Moreover, the work function of the sample can be estimated from KPFM by using $\Phi_{sample} = \Phi_{probe} - e\Delta V_{CPD}$, where $\Phi_{probe}$ is the work function of the probe[33]. Thus, brighter surface potential contrast is associated with lower work function, i.e., work function of 3LG < 2LG < 1LG.

Figure 4d shows the SEM image obtained in a vacuum chamber using the InLens detector to capture backscattered electrons. The IFL/SiC regions appear brightest in the SEM image, which is consistent with the contrast of SEM image for the epitaxial graphene nanoribbon sample (as shown in Fig. 3e and Supplementary Fig. S4b). SEM image of IFL/SiC regions with the brightest contrast has been consistently observed in all the samples presented in this paper and in all the other samples that we have imaged, indicating that IFL/SiC has the lowest work function. 1LG appears the darkest and 2LG has contrast in between the IFL/SiC region and 1LG, indicating work function of IFL/SiC < 2LG < 1LG. Finally, during the SEM imaging, the graphene surface becomes heavily charged by the electron beam and is also exposed to hydrocarbon contamination, both causing deterioration of the image resolution. The dark spots that are tens of nanometers in size, which only appear in the SEM image that was obtained last, are likely to be the hydrocarbon contamination. In contrast to that, CLSM is non-invasive and thus does not influence the imaging quality over time.

**DISCUSSION**

In this paper, we demonstrate CLSM, a fast and non-destructive characterization method for optical imaging of graphene, which produces images of optical intensity and height in ambient air, without any prior sample preparation. The CLSM distinguishes features such as thickness inhomogeneity, folds, tearing, and nanoribbons of graphene on various substrates with high spatial resolution (150 nm).

For graphene on Si/SiO$_2$ substrate, CLSM images show excellent correlation to OM, Raman spectroscopy and AFM height mapping, where the latter two can be used to calibrate the CLSM intensity and height to directly determine graphene layer thickness. For graphene on SiC substrate, the measured reflected intensity from 1LG is ≈ 3% higher than that from adjacent IFL regions and the reflected intensity from 2LG is ≈ 2% higher than the 1LG (see Supplementary Section 6). Through direct correlation to the results from Raman spectroscopy, SEM, and scanning probe microscope including AFM, C-AFM, KPFM, CLSM imaging reveals that epitaxial graphene starts to form at the edges of SiC terraces as parallel graphene nanoribbons. The nanoribbons then merge into a continuous, uniform monolayer graphene under proper processing conditions. Micrometer-sized, bilayer and few layer graphene patches are found as high contrast regions in the CLSM images of overgrown samples. Compared to the complementary methods used in this paper, CLSM not only has a much higher throughput for detecting such regions, it is also insusceptible to surface contamination or surface charging, which will strongly affect the resolution and even the contrast of other imaging techniques such as KPFM or SEM. Although Raman spectroscopy also has advantages as a non-intrusive optical method, and the present micro Raman systems have laser spot sizes of around 800 nm, CLSM can produce a map with higher resolution and on a significantly larger scale, in a much shorter time frame.

We propose that high resolution CLSM images can provide inspection of wafer-scale graphene, selection of material and locations for more efficient fabrication (Supplementary Figs. S9 and S10), as well as analysis of device quality and failure modes (Supplementary Fig. S11). CLSM will be particularly valuable for characterization of 2D materials, which have atomic thickness and are susceptible to surface contamination or surface charging, but have different reflectivity.

**METHODS**

CLSM was performed using an Olympus LEXT OLS4100 system[34] fitted with 5×, 10×, 20×, 50× and 100× objectives (numerical apertures: 0.15, 0.30, 0.60, 0.95 and 0.95, respectively) and with up to 8× further optical zoom. This enables the CLSM to image areas with field of view ranging from 2,560 μm to 16 μm, which translates to total magnification range from 108× to 17,280×. The system employs a 405 nm wavelength violet semiconductor laser, which is scanned in the X-Y directions by an electromagnetic micro-electro-mechanical systems (MEMS) scanner and a high-precision Galvano mirror, and a photomultiplier to capture the reflected light and generate images up to 4096×4096 pixels with horizontal spatial resolution of 150 nm (Supplementary). The confocal optical setup only allows the reflected light that is in-focus to pass through the circular confocal pinhole, thus eliminating flare from out-of-focus regions, but resulting in a very shallow depth of field. To increase the focus resolving capability, a series of images along Z-axis are taken around the median focus height, with separations as small as 60 nm. For each pixel, an ideal *Intensity-Z* curve is calculated to fit the intensities in these images and extract the maximum value, which in turn is used to create a final 2D intensity image. The system is operated in ambient air and does not require any sample preparation for clean samples.

AFM was performed in tapping-mode in air using a Bruker Dimension FastScan SPM[34]. In this mode, the probe oscillates at its fundamental resonance ($f_0$) and a feedback loop tracks the surface of the sample by adjusting the Z-piezo height to maintain a constant amplitude of the cantilever oscillation. The phase of the cantilever oscillation is also compared to the sine wave driving the cantilever oscillation, and thus, AFM achieves simultaneous mapping of the topography and tapping phase, which is a measure of the energy dissipation between the probe and sample, thus encompassing variations in adhesion, composition, friction, viscoelasticity and other mechanical properties of the sample[41].

C-AFM was performed using a Bruker Dimension Icon SPM[34] by raster scanning a Pt probe across the sample surface. The C-AFM scans were performed with 250 mV bias voltage applied to the sample and the resulting current flowing through the probe at each pixel of the scan area was measured by a current amplifier. Epitaxial graphene's high electrical conductivity and good adhesion allow precise mapping of the nanostructures by C-AFM unless they are isolated by non-conducting SiC or interfacial layer (IFL) carbon.

KPFM was performed by means of frequency modulation (FM) using a Bruker Dimension Icon SPM[34]. During FM-KPFM, the surface of the sample is tracked and measured using the AFM feedback method described above. Additionally, a low frequency ($f_{mod}$), AC voltage ($V_{AC}$) is applied to the electrically conductive probe, which shifts the cantilever resonance due to the electrostatic attraction/repulsion and thus produces side lobes at $f_0 \pm f_{mod}$. When the FM-KPFM feedback loop applies an additional DC voltage to the probe ($V_{DC}$), the amplitude of the side lobes is proportional to the difference between $V_{DC}$ and the surface potential of the sample (also referred to as the contact potential difference, $V_{CPD}$). The surface potential is determined by the $V_{DC}$ minimizing the side lobes, i.e., when potential of the probe is equal to the potential of the sample. The surface potential map is obtained by recording $V_{DC}$ pixel by pixel. The surface potential values of the sample can be converted to a work function using, $\Phi_{sample} = \Phi_{probe} - e\Delta V_{CPD}$, provided the work function of the probe ($\Phi_{probe}$) is known. For further details see Ref. [42].

Raman spectra of graphene on Si/SiO$_2$ were acquired under ambient conditions with 514.5 nm excitation (Renishaw InVia)[34], which is focused to an approximately 1 μm spot on the sample through a 50× objective (0.75 NA). The Raman spectra and mapping of epitaxial graphene on SiC were acquired under ambient conditions with 532 nm excitation (Renishaw InVia)[34], which is focused to an approximately 0.8 μm spot on the samples through a 100× objective (0.85 NA). Raman maps were performed by raster scanning the laser with a step size of 100 nm and collecting the spectra with an exposure time of 1 second for each point, 1800 l/mm grating and high confocality (20 μm slit opening). Raman maps of the G- and 2D-peaks area, intensity, width and shift were generated from fitting the data.

**Acknowledgement**

V.P., C.M. and O.K. would like to acknowledge funding from Graphene Flagship, GRACE and NMS. Work done by Y.Y. was supported by federal grant #70NANB12H185. Work done by V.P. at NIST was supported by federal grant. A.F.R would like to thank the National Research Council's

Research Associateship Program for the opportunity. The work of C.-I.L at NIST was made possible by arrangement with Prof. C.-T. Liang of National Taiwan University. We would like to thank Dr. Darwin Reyes-Hernandez for beneficial discussion about confocal microscopy. We would like to thank Dr. T. Shen to transfer the CVD graphene on Si/SiO$_2$ substrate.

**Author Contributions**

V.P., Y.Y. and R.E.E. conceived and designed the experiments. V.P., Y.Y., G.C., J.H., M.K., C.-I.L., A.F.R., C.M., A.R.H.W and R.E.E performed the experiments. Y.Y., J.H. and R.E.E. produced the samples. Y.Y. and C.-I.L. fabricated the devices. Y.Y. and V.P. performed image processing and data analysis. The manuscript was written through contributions of all authors.

**Competing financial interest**

The authors declare no competing financial interest.

**Materials & Correspondence**

*Vishal Panchal, Email: vishal.panchal@npl.co.uk, ORCID: 0000-0003-3954-8535

*Randolph E. Elmquist, Email: randolph.elmquist@nist.gov

**Additional Information**

   **Supplementary Information** is available and includes (1) CLSM characterization of exfoliated graphene on Si/SiO$_2$ substrate, (2) Conventional optical and CLSM intensity imaging of epitaxial graphene nanoribbons, (3) Estimation of lateral resolution for CLSM image, (4) Raman maps for epitaxial graphene nanoribbons, (5) Evolution of the optical contrast features from incomplete EG to continuous EG, (6) Notes on the reflected intensity from IFL, 1LG and 2LG, (7) Large area EG characterization by CLSM image stitching, and (8) Device inspection by CLSM.

# Confocal laser scanning microscopy: A tool for rapid optical characterization of 2D materials

## *(Supplementary Information)*

Vishal Panchal[1,2], Yanfei Yang[2,3], Guangjun Cheng[2], Jiuning Hu[2], Mattias Kruskopf[2], Chieh-I Liu[2,4], Albert F. Rigosi[2], Christos Melios[1], Angela R. Hight Walker[2], David B. Newell[2], Olga Kazakova[1], and Randolph E. Elmquist[2]

[1]National Physical Laboratory, Hampton Road, Teddington, TW11 0LW, UK

[2]National Institute of Standards and Technology, Gaithersburg, MD 20899, USA

[3]Joint Quantum Institute, University of Maryland, College Park, MD 20742, USA

[4]National Taiwan University, Taipei, 10617, Taiwan

**Contents**

## 1. CLSM characterization of exfoliated graphene on Si/SiO₂ substrate

Figures S1a and S1b are the conventional optical and CLSM height images, respectively, of the same exfoliated graphene flake as shown in Fig. 1 in the main text. The inset of Fig. S1b is the AFM topography image of region marked by yellow box in Fig. S1b. Figure S1c is the Raman spectra for all nine spots marked in Fig. 1b in main text, where the G-peak height increases with the layer number. Figure S1d shows the CLSM intensity and height profiles for the cross section indicated by the red dashed line in Fig. S1b, showing that both properties increase linearly with the layer number. Figure S1e compares the CLSM height profile to the AFM height profile for the same cross section. While the AFM measurement of the first graphene layer is strongly affected by the gap between the substrate and the graphene flake, the CLSM intensity and height measurements are not so susceptible to this problem.

In Figure S2, we applied linear fitting for both CLSM and AFM height measurement as a function of the layer number and obtained the following equations:

$$T_{g-CLSM} = 3.6767 \times H_{CLSM} + 0.8986 \tag{1}$$

$$T_{g-AFM} = 0.4800 \times H_{AFM} + 0.7728 \tag{2}$$

Where $T_g$ is the thickness of graphene and $H$ is the height measured by either CLSM or AFM. The data of graphene height plotted in Fig. S2 are extracted from Figs. 1b-1d in the main text using the "statistical quantities" tool in Gwyddion[1,2], corresponding to the 7 spots listed in Table S1. From the above equations, we can calibrate the CLSM height measurement as

$$T_{g-CLSM-cal} = \frac{0.4800}{3.6767} \times T_{g-CLSM} + \left[ 0.7728 - \left( \frac{0.4800}{3.6767} \times 0.8986 \right) \right] \tag{3}$$

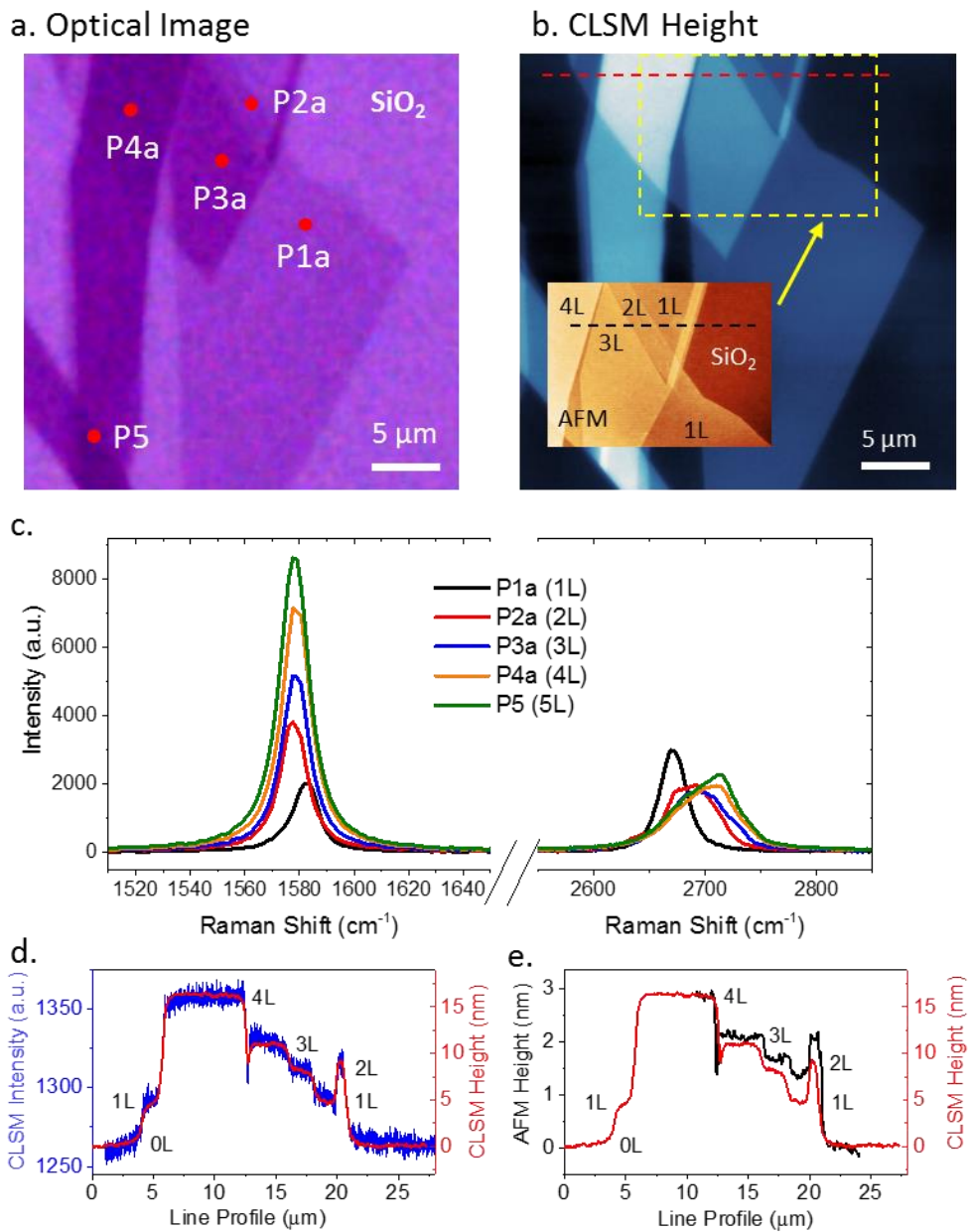

**Figure S1.** (a) Optical image of the same exfoliated graphene flake as discussed in Fig. 1 in the main text. (b) CLSM height image of the same flake as in (a). (c) The Raman spectra of 1-5 layer graphene marked as in Fig. 1b in the main text showing the evolution of G- and 2D-peaks with the layer number. (d) The CLSM intensity (blue) and CLSM height (red) profiles along the red dashed line shown in (b). (e) The AFM height (black) and CLSM height (red) line profiles along the black and red dashed lines shown (b). Raman data was acquired with 514.5 nm excitation.

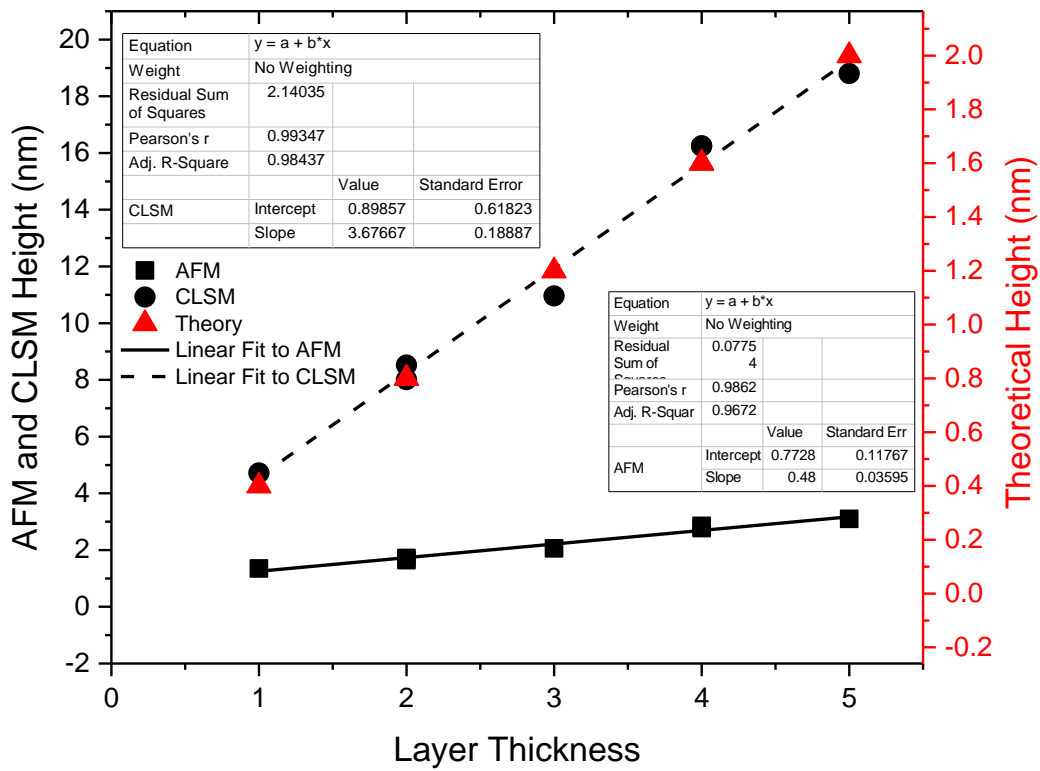

**Figure S2.** Black solid line is linear fitting to the AFM height data points (black squares) and black dashed line is linear fitting to the CLSM height data points (black circles), both as a function of the layer numbers.

**Table S1.** Summary of the AFM height and CLSM height data used in Fig. 1 in the main text.

| Points | AFM height (nm) | Error (nm) | CLSM height (nm) | Error (nm) |
|--------|-----------------|------------|------------------|------------|
| P1b | 1.35 | 0.10 | 4.72 | 0.11 |
| P2a | 1.64 | 0.11 | 8.01 | 0.20 |
| P2b | 1.72 | 0.13 | 8.52 | 0.13 |
| P3a | 2.06 | 0.10 | 10.96 | 0.11 |
| P4a | 2.85 | 0.16 | 16.24 | 0.12 |
| P4b | 2.84 | 0.11 | 16.25 | 0.16 |
| P5 | 3.11 | 0.11 | 18.55 | 0.12 |

## 2. Conventional optical and CLSM intensity images of epitaxial graphene nanoribbons

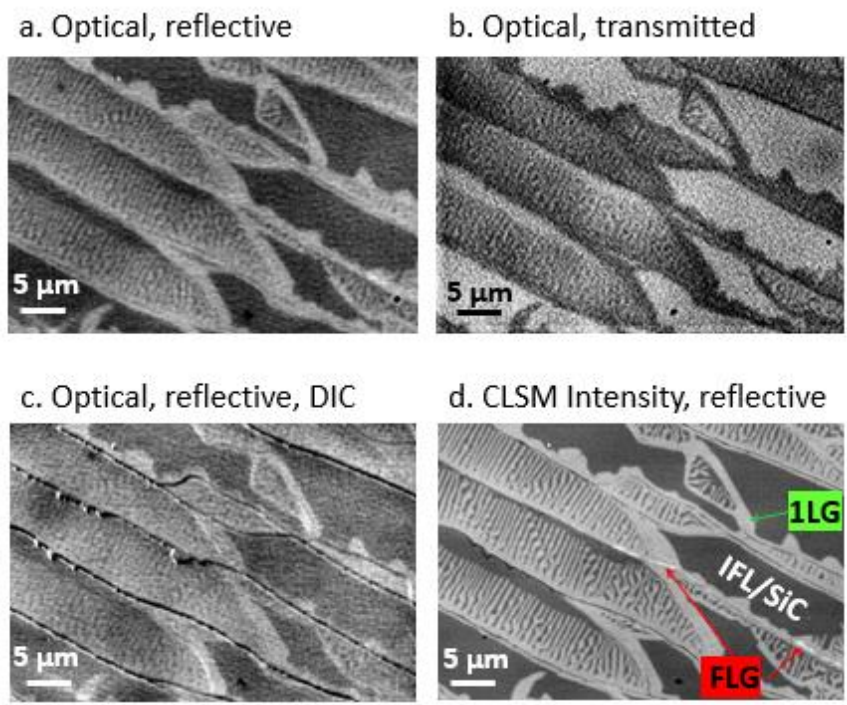

**Figure S3.** Optical images obtained in (a) reflection, (b) transmission, (c) reflection with differential interference contrast (DIC) and CLSM intensity in reflection modes.

The conventional optical images of epitaxial graphene on SiC in Fig. S3 were obtained with a Nikon Eclipse L200N optical microscope[2], which is capable of operating in the reflection, transmission or differential interference contrast (DIC) mode using white light. The brighter regions of the reflective optical image are associated to graphene which is partially covering the surface[3], and darker regions are insulating interfacial layer (IFL) or bare SiC, as verified later with a host of scanning probe microscopy and Raman techniques. Figure S3b shows the transmission image of the same area as Fig. S3a, where the contrast is now inverted. Imaging in the DIC mode produces a 3-dimensional visualization of the surface morphology with certain sacrifice of the graphene contrast. Fig. S3d is the CLSM image of the same area as in Figs. S3a-c, which not only shows a much higher special resolution, but also clearly reveals thin stripes (labeled by red arrows) and patches of higher reflectivity along the step edges, indicating few-layer graphene (FLG).

### 3. Estimation of lateral resolution for CLSM image

Figure S4a-b shows the analysis of a CLSM and SEM image containing dense forest of 2D graphene nanoribbons by using the image processing program ImageJ[2]. The red curve in Fig. S4c is the averaged profile crossing the graphene nanoribbons marked by the red rectangular box, compared to the averaged profile (blue curve) from the same region of the SEM image. For the nineteen dips clearly seen from the SEM profile, seventeen corresponding peaks (counted from right) can be distinguished from the CLSM profile for the graphene nanoribbons with width varying approximately from 120 nm to 230 nm.

We further estimated the lateral resolution of the CLSM image in Fig. S4a by analyzing the edge spread function (ESF). The blue points in Fig. S4d-e are the averaged profile across the edges marked by two red filled boxes in Fig. S4a. An integrated Gauss function is used to fit the averaged profile, with the high plateau on the left defined as 100% brightness and the low plateau on the right defined as 0% brightness. The lateral resolution is estimated by calculating the edge width between two reference points with 20% and 80% of brightness (Fit 20/80). The inset label in Fig. S4d-e shows that the lateral resolution is approximately 149 nm and 161 nm for the red filled boxes marked 1 and 2 in Fig. S4a, respectively. The lateral resolution will be affected by factors such as the contrast level and materials, and therefore varies from sample to sample, or from region to region on the same sample.

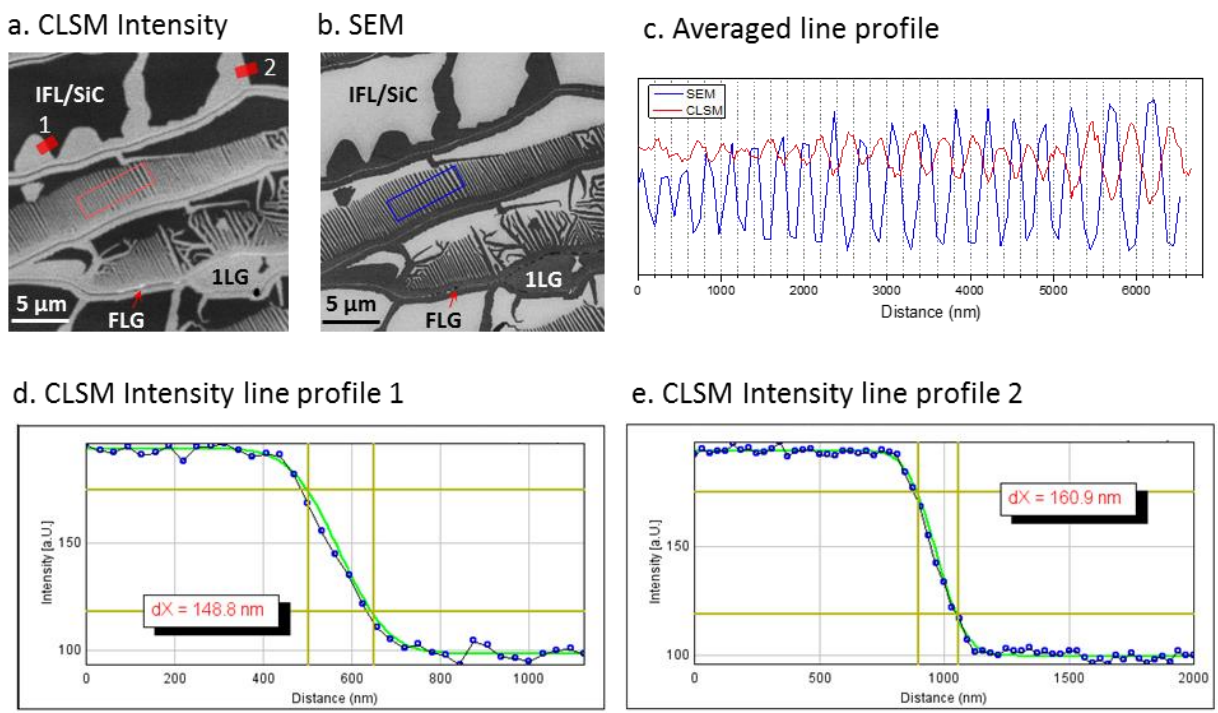

**Figure S4.** Analysis of the lateral resolution of the CLSM and SEM EG images by ImageJ. (a) CLSM and (b) SEM images of graphene nanoribbons on SiC. (c) Averaged profile for the red and blue rectangular boxes in (a-b). (d) and (e) CLSM intensity profiles at the edges of graphene and IFL/SiC indicated by the red filled boxes marked 1 and 2 in (a), with Gauss simulation (green line).

Figure S5 shows the CLSM image of an epitaxial graphene sample containing regions of IFL, 1LG 2LG, and 3LG. The lateral resolution values estimated from the CLSM image in Figure S5 show that the 20/80% of ESF for the edges denoted in the table on the right vary from approximately 97.3 nm to 185.8 nm. Based on analysis of more than 10 CLSM images from different samples, we estimate that the lateral resolution of our CLSM images is approximately 150 nm.

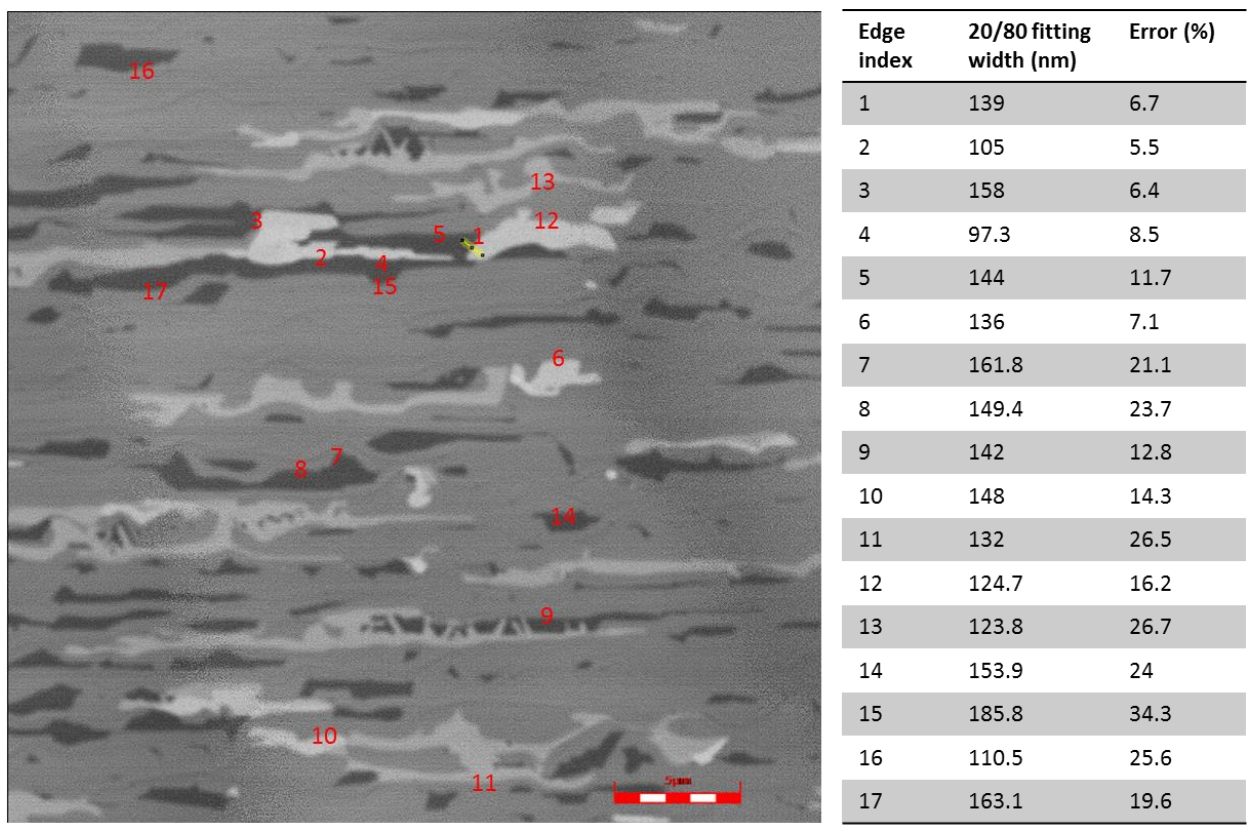

| Edge index | 20/80 fitting width (nm) | Error (%) |
|---|---|---|
| 1 | 139 | 6.7 |
| 2 | 105 | 5.5 |
| 3 | 158 | 6.4 |
| 4 | 97.3 | 8.5 |
| 5 | 144 | 11.7 |
| 6 | 136 | 7.1 |
| 7 | 161.8 | 21.1 |
| 8 | 149.4 | 23.7 |
| 9 | 142 | 12.8 |
| 10 | 148 | 14.3 |
| 11 | 132 | 26.5 |
| 12 | 124.7 | 16.2 |
| 13 | 123.8 | 26.7 |
| 14 | 153.9 | 24 |
| 15 | 185.8 | 34.3 |
| 16 | 110.5 | 25.6 |
| 17 | 163.1 | 19.6 |

**Figure S5.** Left: CLSM image of EG sample covered by dominant single layer graphene (1LG). The lowest brightness indicates IFL region. Higher brightness corresponds to thicker graphene layer. Right: A table with edge width calculated from "Fit 20/80" algorithm.

## 4. Raman maps for epitaxial graphene nanoribbons

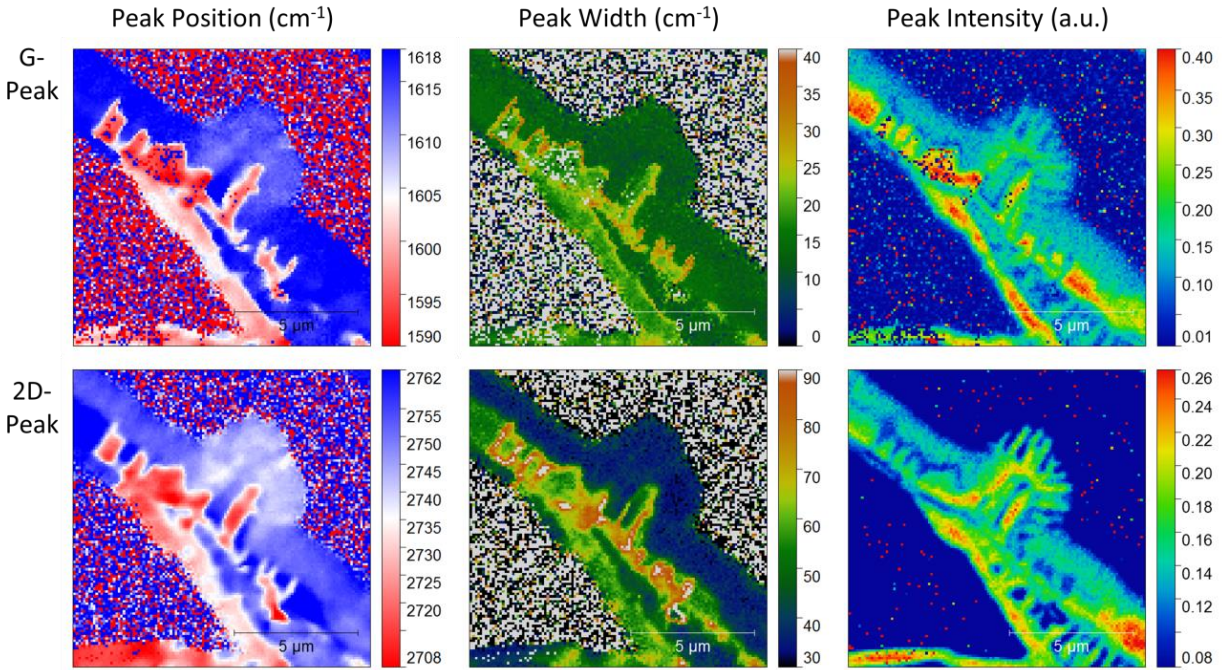

**Figure S6.** Raman maps of the peak position, width and intensity for the G- and 2D-peaks for the epitaxial graphene nanoribbons area presented Fig. 3 in the main text. Raman data was acquired with 532 nm excitation.

## 5. Evolution of the optical contrast features from incomplete EG to continuous EG

Figure S7a is a reflective optical image of a face-to-graphite sample obtained by Nikon Eclipse L200N optical microscope[2] with a 50× objective, showing incomplete single layer graphene (1LG) coverage. The graphene nanoribbons merged into continuous graphene in a succeeding growth, as shown in Fig. S2b. The conspicuous contrast from the interfacial layer regions (the darker contrast in Fig. S2a) disappeared in Fig. S2b. Instead, only narrow lines of higher brightness are seen after the second growth along the step edges, indicating few layer graphene, as confirmed by Raman spectroscopy. Figure S7c and S7d are cropped from Fig. S7a and S7b, respectively, showing the same region where a Raman map (Fig. S7e) has been generated after the second growth. The spectrum from a spot on the terrace, marked by a green circle in Fig. S7e, show a symmetric 2D-peak (the green curve in Fig. S7f) that can be fit by a single Lorentzian (the black dashed line in Fig. S7f), confirming the existence of 1LG. The spectrum from the spot at the step edge, marked by a red circle in Fig. S7e, shows a much wider asymmetric 2D-peak (the red curve in Fig. S7f), indicating few layer graphene.

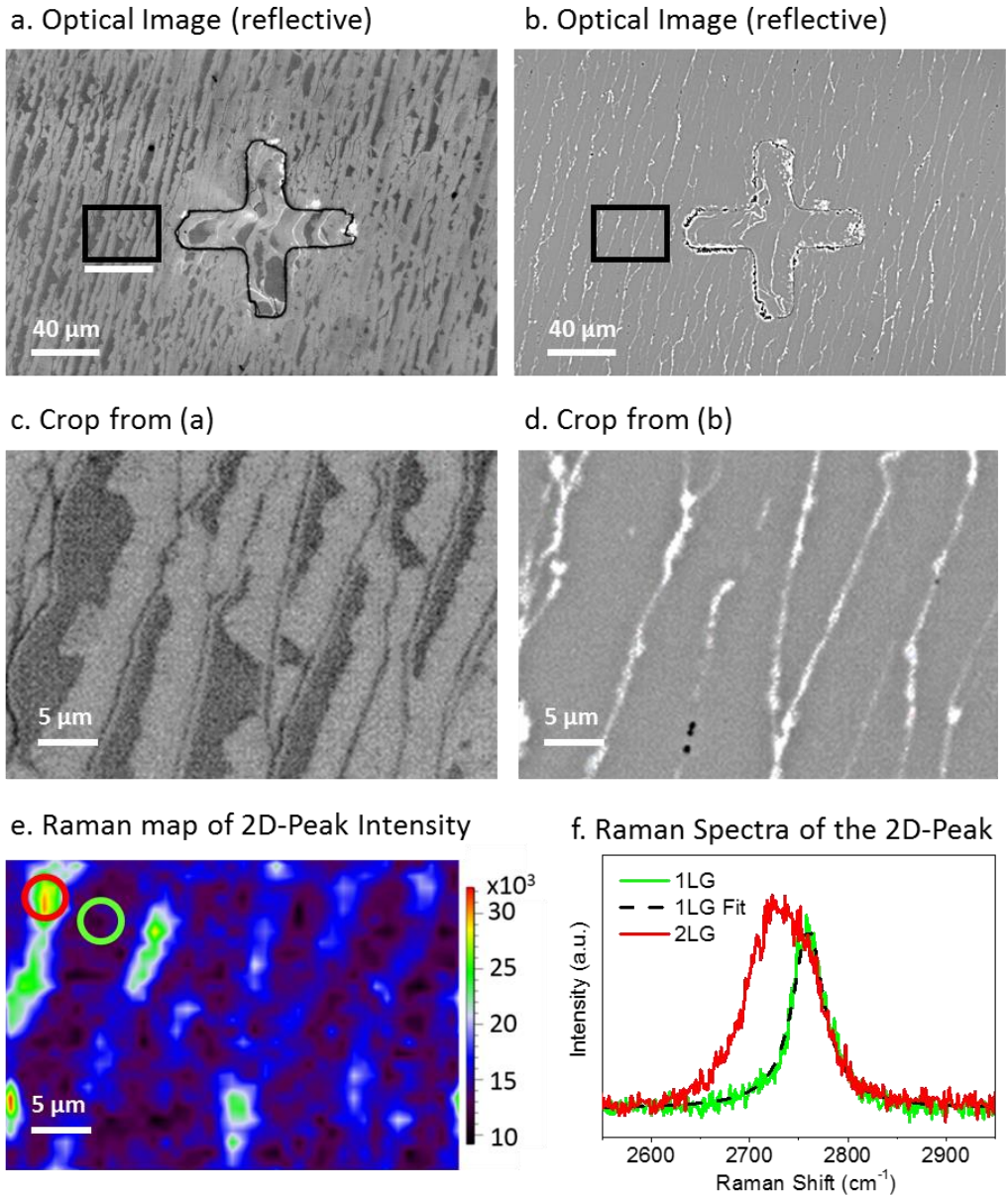

a. Optical Image (reflective)

b. Optical Image (reflective)

c. Crop from (a)

d. Crop from (b)

e. Raman map of 2D-Peak Intensity

f. Raman Spectra of the 2D-Peak

**Figure S7.** (a) Conventional reflective optical image of a face-to-graphite sample with partial graphene coverage. (b) Conventional reflective optical image of the same sample after a succeeding growth showing continuous background with narrow lines of higher brightness. (c) and (d) Cropped images of the region marked by the black boxes in (a) and (b) respectively. (e) 2D-Peak intensity Raman map of the same region as in (d) after the second growth. (f) Raman spectra for 1LG and 2LG from the spots marked by green and red circles in (e), respectively. The black dot line is the Lorentzian fitting of the green curve. Raman data was acquired with 514.5 nm excitation.

## 6. Notes on the reflected intensity from IFL, 1LG and 2LG

We have found that the sharpness level (Fig. S8) in the advanced settings for the CLSM will strongly affect the reflected intensity due to the backstage algorithm. As suggested by the Olympus[2] specialist, we turned off the contrast and sharpness enhancement when estimating the ratio of reflected intensity from IFL, 1LG and 2LG. We have universally observed that the reflected intensity from 1LG is ~3 % higher than that from IFL region, and the reflected intensity from 2LG is ~2 % higher than that from 1LG.

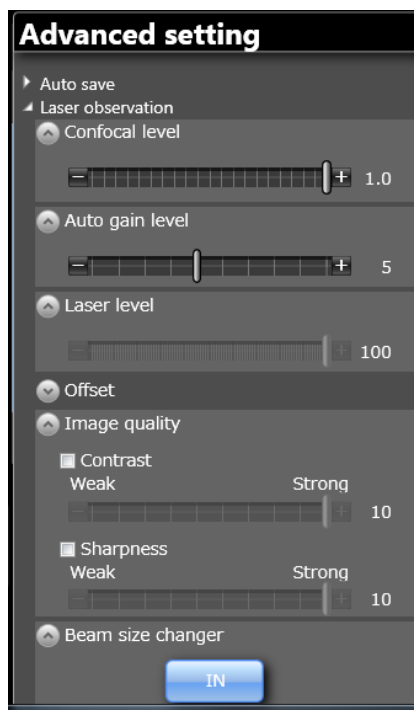

**Figure S8.** CLSM advanced setting used for estimation of the change of reflected intensity from IFL, 1LG and 2LG.

## 7. Large area EG characterization by CLSM image stitching

Since the graphene nanoribbons as well as the 2LG and FLG patches are usually submicron sized, a single CLSM scan by 20× objective and higher magnification cannot distinguish such features properly and are not suitable for the characterization of EG region larger than hundreds of micrometers. Wafer-scale EG can be characterized by stitching arrays of CLSM images scanned by 50× or 100× objective as shown in Fig. S9 and Fig. S10. Figure S9a shows a high resolution CLSM image (produced from 64 CLSM scans by digital stitching) of a homogeneous monolayer graphene area (463 μm by 463 μm) that includes less than 1% of multilayer graphene (the irregular brighter patches). Figures S9b, S9c, and S9d are the three zoomed-in grid CLSM images for locations marked by red boxes 1, 2 and 3 in Fig. S9a. Hall bar devices of 400 μm width fabricated from such graphene can maintain the quantum Hall effect with precise metrological accuracy up to 4 K.

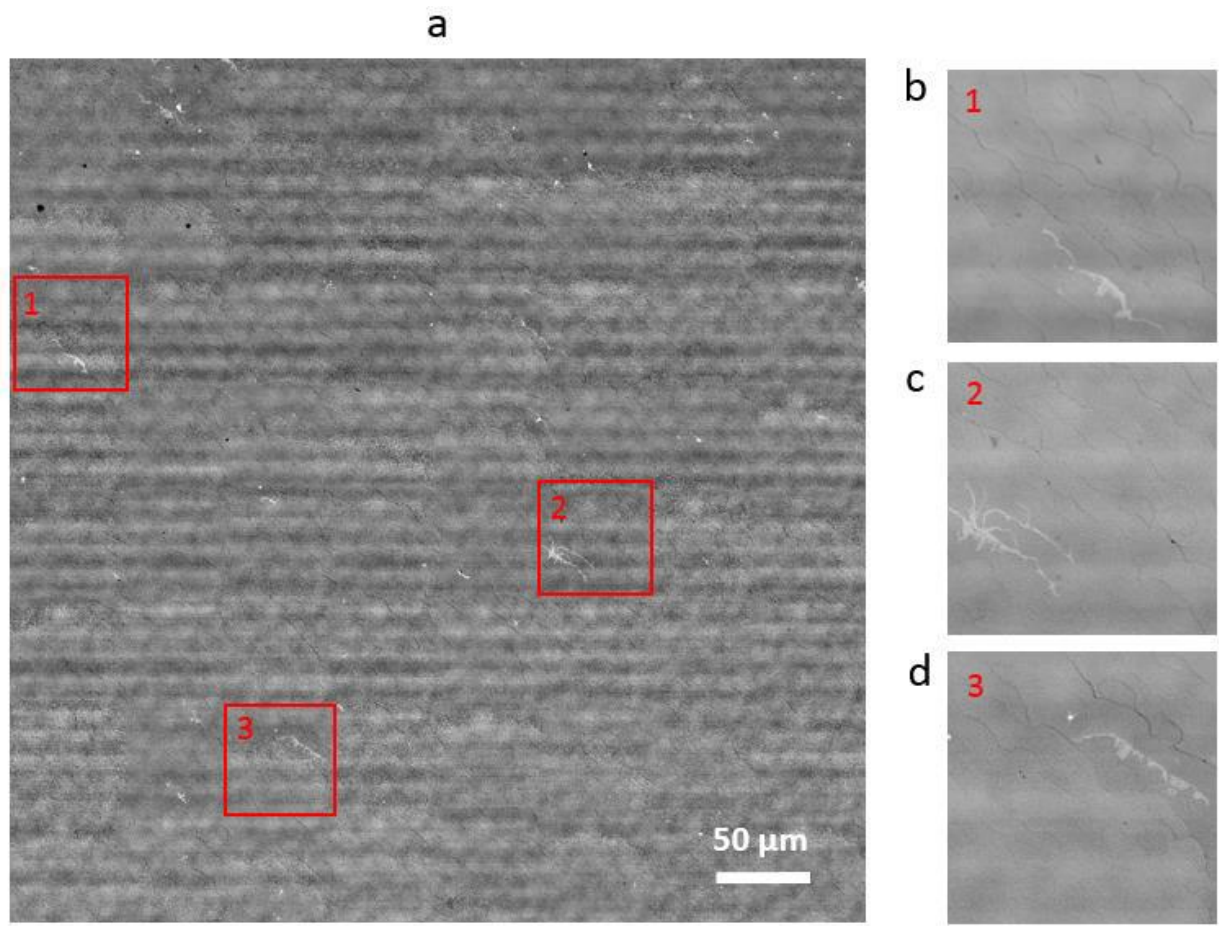

**Figure S9.** (a) Stitched CLSM image of a highly uniform area of monolayer graphene with only few bilayer patches as shown in the right panels. (b)-(d) Zoom-in of the region indicated by the red boxes 1-3 in (a), respectively.

Fig. S10 shows a composite image produced in ~20 minutes from 16 CSLM scans by digital stitching. The black strip that appears in the lower region of this image is the edge of the sample. The fiducial mark (V20) is used for sample identification, and is etched into the SiC before EG is grown. face-to-graphite growth usually produce very thick graphene layers[4] close to the edge of the sample (region 3 with much higher brightness in Fig. S10). About few hundreds of micrometers away from the edge, bilayer and few layer patches decrease dramatically in region 2. Continuous

EG with less than 1% of bilayer or few layer patches in region 1 is suitable for fabrication of quantum Hall resistance standards[4].

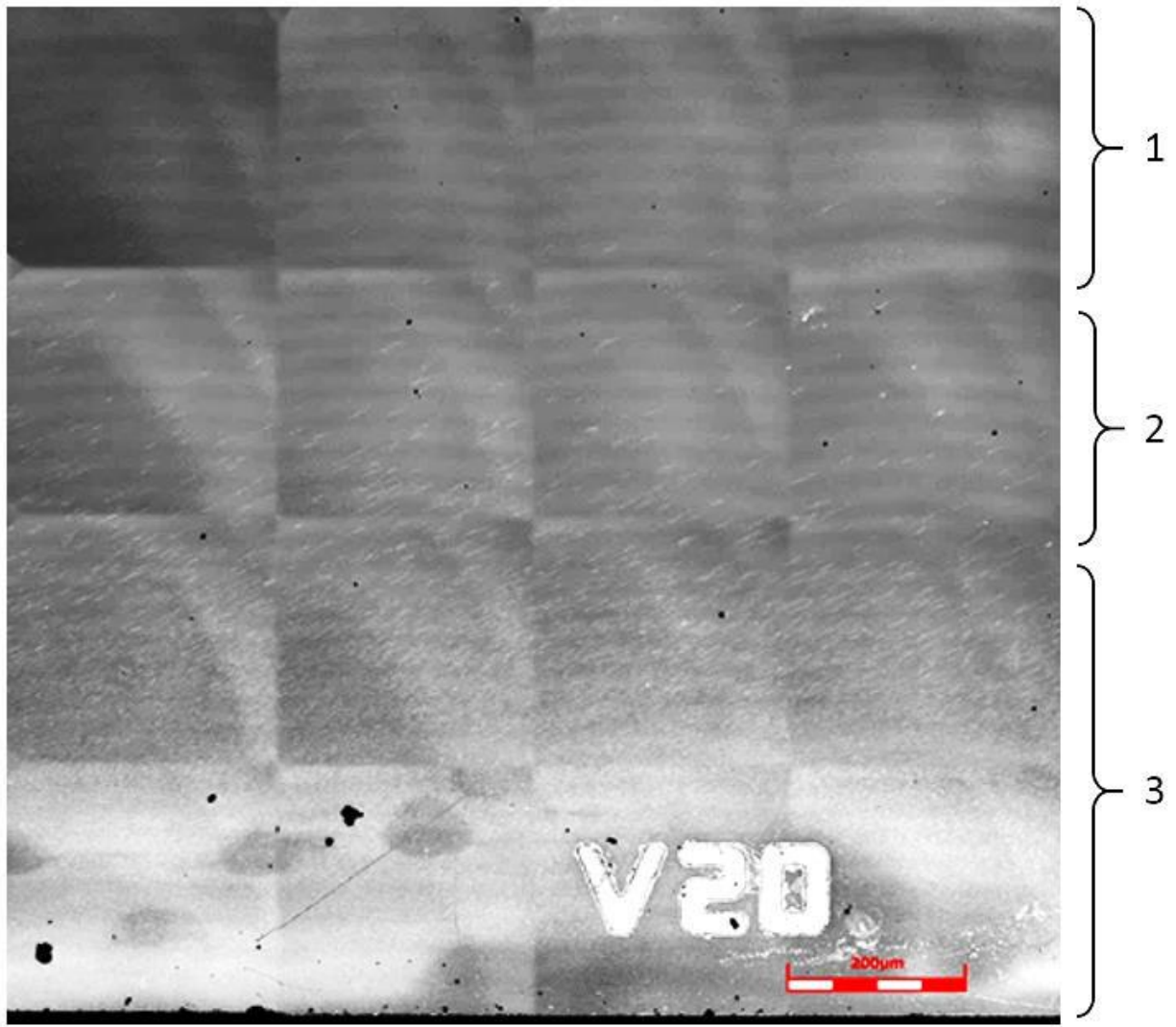

**Figure S10.** Stitched CLSM image of an area near the edge of a primarily monolayer EG sample, where thicker graphene patches (bright contrast) can only be seen near the edge. Region 1 is covered by uniform and continuous 1LG. Region 2 shows increasing bilayer and few layer patches (with higher brightness). Region 3 is covered by very thick graphene.

8. **Device inspection by CLSM**

Charge carrier mobility of graphene is an important electronic property that is usually measured using the Hall effect. However, the mobility of epitaxial graphene is strongly affected by its carrier density. To compare the quality of two graphene devices, one needs to compare the curves of mobility as a function of carrier density obtained at low temperature, as shown in Figure S11. Here we correlate the mobility characteristic curves to the CLSM images of corresponding devices. The CLSM image (left inset in Fig. S11) of the high mobility device (red data in Fig. S11) shows almost complete graphene coverage with less than 10% of bilayer or interfacial layer inclusions. The CLSM image (right inset in Fig. S11) of the low mobility device (black data in Fig. S11) shows large portion of interfacial layer.

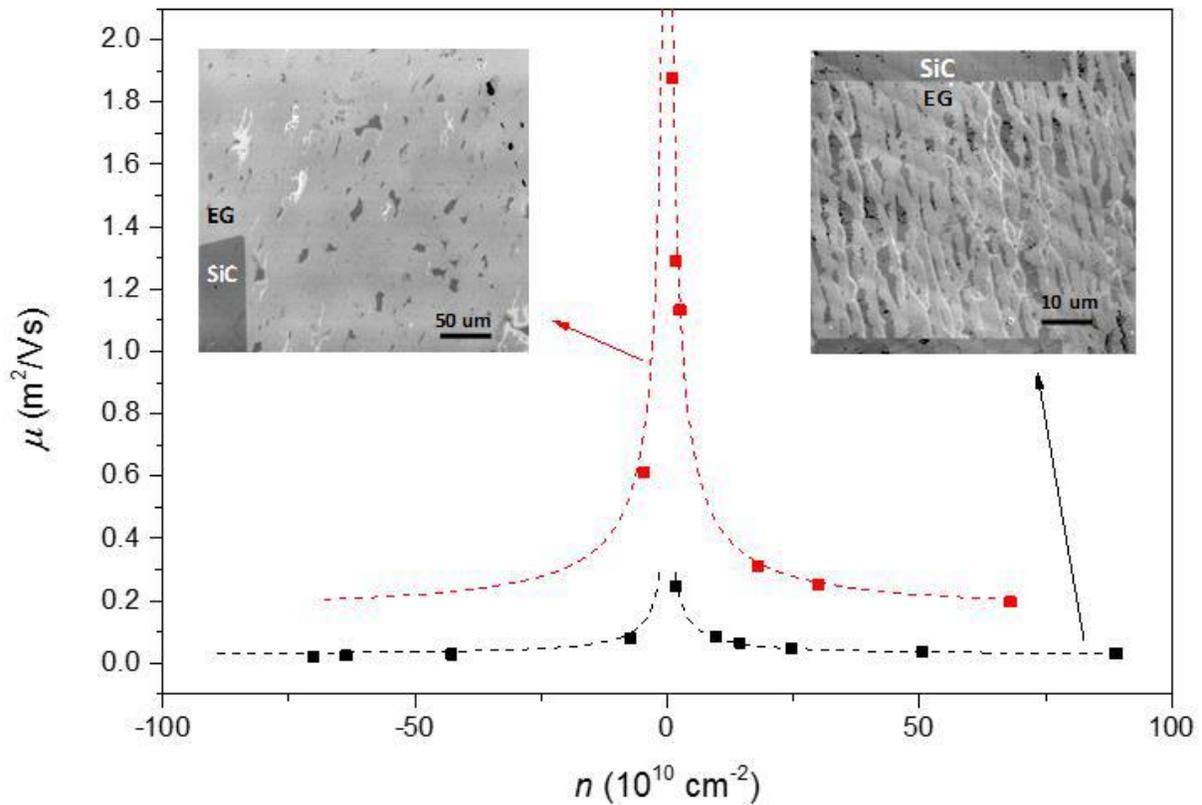

**Figure S11.** Mobility of EG devices as a function of carrier density. The overall mobility of a more uniform sample (data shown in red, CLSM image in left inset) is much higher than that of another

device (data shown in black, CLSM image in right inset) made from graphene area with incomplete graphene and nanoribbons.